# Models of Short-Time Qubit Decoherence

Dmitry Solenov and Vladimir Privman

Center for Quantum Device Technology
Clarkson University, Potsdam, New York 13699–5721, USA

Electronic addresses: solenov@clarkson.edu, privman@clarkson.edu

## ABSTRACT

We compare approaches to evaluation of decoherence at low temperatures in two-state quantum systems weakly coupled to the environment. By analyzing an exactly solvable model, we demonstrate that a non-Markovian approximation scheme yields good quality estimates of the reduced density matrix for time scales appropriate for evaluation of quantum computing designs.

**Keywords:** decoherence, environment, quantum computing, qubit, relaxation, spin, thermalization

## 1. INTRODUCTION

The possibility of realizing controllable two-level quantum systems (qubits) has been of central importance in the field of quantum information processing. Spin degrees of freedom of individual electrons[1,2] and nuclei,[3] as well as superconducting circuits[4-7] and semiconductor charge qubits[8] have been studied extensively. Description of decoherence for such systems has been recognized as crucial for evaluation of feasibility of fault-tolerant quantum computing.[9] Often the interaction between the system and environment can be considered quite weak. However, in most cases it is impossible to obtain an exact solution to the evolution of the system's reduced density matrix with the environmental modes traced over. This has led to the development of various approximation techniques for estimating the system dynamics on different time scales.[10-17] These include Markovian-type approximations,[10,11] which describe the exponential decay toward the thermalized, decohered state. Recently, it has been argued[13,18,19] that this approach is valid only for relatively long times and does not represent the dynamics for short times appropriate for the quantum-computing gate functions. New short-time approximation schemes have been advanced.[13,16]

The total Hamiltonian of the system is represented as

$$H = H_S + H_B + H_{int} .\tag{1}$$

It is common to model the environment by a bath of modes, with Hamiltonians $M_k$,

$$H_B = \sum_k M_k .\tag{2}$$

The interaction part is defined as

$$H_{int} = \Lambda X = \Lambda \sum_k X_k ,\tag{3}$$

where $\Lambda$ is a Hermitian operator in the system space, coupled to the bath operators $X_k$. For the standard choice of the bosonic bath modes and linear coupling,[20-25] we have

$$M_k = \omega_k b_k^\dagger b_k \tag{4}$$

where the ground state energy is shifted to zero, and

$$X_k = g_k b_k^\dagger + g_k^* b_k .\tag{5}$$

Here and throughout the paper we use $\hbar = 1$. Irreversibility arises as a result of the system's interaction with an infinite number of the bath modes, in the continuum limit.[26] The system is described by a reduced density matrix, obtained by tracing the total density matrix over all bath modes,

$$\rho(t) = Tr_B[\rho_{total}(t)] .\tag{6}$$

It has been argued[13] that short-time decoherence processes are crucial for evaluation of fault-tolerance in quantum computing. The consideration of small times allows the use of several approximations. The simplest one is the expansion of the evolution operator in powers of time, provided the expansion is considered up to the second order (SO) only.[14] An improved approach[13] is the approximation that is based on the exact solvability of the adiabatic dynamics[15] (AD). Recently, a new approximation[16] based on a perturbative expansion that gives a non-Markovian (NM) dynamics has been proposed.

In the present work, we consider the exactly solvable model of adiabatic decoherence,[15] for which the AD approximation[13] is exact, and calculate the SO and NM approximations at low temperatures, in order to evaluate decoherence of a single qubit, coupled linearly to an Ohmic bath[25] of modes. The resulting approximations are then compared by calculating a recently introduced[27] additive measure of decoherence.

It has been demonstrated[15,28] that for the adiabatic case of no energy exchange with the bath, $[\Lambda, H_S] = 0$, the exact expression for the density matrix elements in the energy basis can be obtained. The absolute values are given by

$$|\rho_{mn}(t)| = |\rho_{mn}^{ideal}(t)| e^{-B^2(t)\delta_{mn}} , \tag{7}$$

and illustrate the process of adiabatic decoherence: the off-diagonal density matrix elements vanish for large times. Here

$$B^2(t) = 8 \int d\omega |g(\omega)|^2 G(\omega) \omega^{-2} \sin^2(\omega t/2) \coth(\omega t/2) , \tag{8}$$

$\rho^{ideal}(t)$ is the density matrix of the isolated system, and $G(\omega)$ represents the density of states of the bath.

We will obtain specific results for a single-qubit system with adiabatic coupling, defined by

$$H_S = \frac{E}{2}\sigma_z , \qquad \Lambda = \sigma_z , \tag{9}$$

where $\sigma_z$ is a Pauli matrix and $E > 0$.

Estimation of the rate of decoherence can be based on calculation of various quantities, such as quantum entropy,[29] fidelity,[30,31] or norm of deviation.[27] We will use the latter measure,

$$D(t) = \max_{\rho(0)} \left( \max_i |\lambda_i| \right) , \tag{10}$$

where $\lambda_i$ denote the eigenvalues of the deviation operator

$$\Delta\rho(t) \equiv \rho(t) - \rho^{ideal}(t) . \tag{11}$$

This measure was shown[27] to be approximately additive for the multi-qubit case.[32]

The technical part of the present work involves the calculation of the reduced density matrix for the above model within the NM approximation, reported in Section 3, while in Section 2 we present general results within the NM approach,[16] that do not depend on the choice of the interaction. Section 4 is devoted to the calculation of the norm-of-deviation measure within the SO, NM and AD approaches and discussion of the results.

## 2. GENERAL FORMULATION

The evolution of the density matrix is given by the Liouville equation

$$i\dot{\rho}_{total}(t) = L\rho_{total}(t) , \tag{12}$$

where the dot denotes the time derivative, and $L\diamond = [H, \diamond]$ is the Liouville superoperator. Equation for the reduced density matrix can be conveniently obtained by introducing the "relevant," $\rho_R(t)$, and "irrelevant," $\rho_I(t)$, parts of the density matrix using the projection superoperators

$$P\diamond = \rho_B Tr_B(\diamond), \qquad Q = 1 - P, \tag{13}$$

where

$$\rho_B = e^{-H_B/kT}/Z, \qquad Z \equiv Tr_B\left(e^{-H_B/kT}\right). \tag{14}$$

Here $\rho_B$ is the thermalized *initial* density matrix of the bath. We define

$$\rho_R(t) \equiv P\rho_{\text{total}}(t) = \rho_B \rho(t), \qquad \rho_I(t) \equiv Q\rho_{\text{total}}(t), \tag{15}$$

The introduction of the initial bath state is done for convenience, to have

$$\rho_R(0) = \rho_{\text{total}}(0) = \rho_B \rho(0). \tag{16}$$

The irrelevant part of the density matrix is initially zero. With these conventions, we can rewrite (12) in the form

$$i\dot{\rho}_R(t) = PL(\rho_R(t) + \rho_I(t)), \tag{17}$$

$$i\dot{\rho}_I(t) = QL(\rho_R(t) + \rho_I(t)). \tag{18}$$

We then substitute integrated equation (18) in (17), to obtain[33]

$$i\dot{\rho}_R(t) = PL\rho_R(t) - i\int_0^t PLe^{-i(t-t')QL}QL\rho_R(t')dt', \tag{19}$$

which turns into an equation for the reduced density matrix, using (13)-(15),

$$i\dot{\rho}(t) = L_S\rho(t) - i\int_0^t Tr_B Le^{-i(t-t')QL}QL\rho(t')\rho_B dt', \tag{20}$$

where we define

$$L_S\diamond = [H_S, \diamond] \quad \text{and} \quad L_0\diamond \equiv [H_S + H_B, \diamond]. \tag{21}$$

The NM-approximation[16] is then

$$i\dot{\rho}(t) = [H_S, \rho(t)] - i\int_0^t Tr_B\left\{[\Lambda X, e^{-i(t-t')L_0}[\Lambda X, \rho_B \rho(t')]]\right\}dt', \tag{22}$$

where the interaction part of the total Hamiltonian enters as $\Lambda X$; see (3). In writing (22), we assumed for convenience that $Tr_B(X\rho_B) = 0$, which can be always obtained by the redefinitions $H_S \to H_S + \Lambda Tr_B(X\rho_B)$, $X \to X - Tr_B(X\rho_B)$. The key approximation in (22) is the assumption that the interaction with the environment is weak, so that only terms up to the second order in it are kept: $e^{iQLt} \to e^{iQL_0 t}$.

We will now derive a formal solution of (22) for the case of a single-qubit (two-state) system, for which we use the spin-½ notation and expand the reduced density matrix in terms of the Pauli matrixes, $\sigma_0 \equiv 1, \sigma_x, \sigma_y, \sigma_z$,

$$\rho(t) = \frac{1}{2} \sum_{\mu=0}^{3} \xi_\mu \sigma_\mu ,  \qquad (23)$$

where $\xi_\mu(t) \equiv Tr_S[\sigma_\mu \rho(t)]$. We have $\xi_0(t) \equiv 1$, while the polarization vector components $\xi_{1,2,3}(t)$ can be shown to satisfy the equations

$$\dot{\xi}_\mu(t) = -\frac{i}{2} \sum_{\nu=0}^{3} \xi_\nu(t) Tr_S\{\sigma_\mu[H_S, \sigma_\nu]\} - \sum_{\nu=0}^{3} \int_0^t I_{\mu\nu}(t') \xi_\nu(t-t') dt' ,$$

$$I_{\mu\nu}(t') = \frac{1}{2} Tr_S\left(\sigma_\mu Tr_B\{[\Lambda X, e^{-iH_0 t'}[\Lambda X, \rho_B \sigma_\nu] e^{iH_0 t'}]\}\right) . \qquad (24)$$

with $H_0 = H_S + H_B$.

Let us now define the interaction-representation time-dependent operators. For an arbitrary operator $\Upsilon$, we have

$$\Upsilon(t) = e^{i(H_S + H_B)t} \Upsilon e^{-i(H_S + H_B)t} . \qquad (25)$$

The kernel $I_{\mu\nu}(t')$ then depends on the correlation function

$$C(t) \equiv Tr_B[X X(t) \rho_B] , \qquad (26)$$

$$2I_{\mu\nu}(t) = C(-t) Tr_S\left[\sigma_\mu \Lambda \Lambda(-t) \sigma_\nu(-t)\right] - C(-t) Tr_S\left[\Lambda \sigma_\mu \Lambda(-t) \sigma_\nu(-t)\right]$$
$$- C(t) Tr_S\left[\sigma_\mu \Lambda \sigma_\nu(-t) \Lambda(-t)\right] + C(t) Tr_S\left[\Lambda \sigma_\mu \sigma_\nu(-t) \Lambda(-t)\right] . \qquad (27)$$

By separating the real and imaginary parts, $C(t) = C_{\text{Re}}(t) + i C_{\text{Im}}(t)$, which satisfy

$$C_{\text{Re}}(t) = C_{\text{Re}}(-t) , \qquad C_{\text{Im}}(t) = -C_{\text{Im}}(-t) , \qquad (28)$$

which follows from the definition of the correlation function, after some algebra one can show that

$$I_{\mu\nu}(t) = C_{\text{Re}}(-t) Tr_S \{\sigma_\nu(-t)[\sigma_\mu, \Lambda]\Lambda(-t)\} . \tag{29}$$

Further simplifications of this kernel require specifying the interaction and will be done in the next section.

In the rest of this section, let us obtain a formal solution for $\rho(t)$ in terms of $\xi_{1,2,3}(t)$; see (23). From now on, we assume that the indices $\mu, \nu$ denote the vector components $1, 2, 3$. Equations (24) can be rearranged in the form

$$\dot{\xi}_\mu(t) = \sum_{\nu=1}^{3} \int_0^t R_{\mu\nu}(t') \xi_\nu(t-t') dt' + k_\mu(t) , \tag{30}$$

where

$$R_{\mu\nu}(t) \equiv -I_{\mu\nu}(t) + \frac{1}{2} i\delta(t) Tr_S \{\sigma_\mu [H_S, \sigma_\nu]\} , \tag{31}$$

$$K_\mu(t) \equiv -I_{\mu 0}(t) + \frac{1}{2} i\delta(t) Tr_S \{\sigma_\mu [H_S, \sigma_0]\} . \tag{32}$$

The Laplace transform of $\xi_\nu(t)$,

$$\xi_\nu(s) = \int_0^\infty e^{-st} \xi_\nu(t) dt , \tag{33}$$

then satisfies

$$\vec{\xi}(s) = [\vec{R}(s) - s]^{-1} [-\vec{\xi}(t=0) - \vec{K}(s)] , \tag{34}$$

where we use the vector/dyadic notation. The general solution for the NM approximation is given by the inverse Laplace transform of (34).

## 3. ADIABATIC QUBIT-BATH INTERACTION

In this section we consider the adiabatic spin-boson model defined by (4), (5) and (9). The expression for the correlation function becomes

$$C(t) = \sum_{n,m} Tr_B \{(g_n b_n^\dagger + g_n^* b_n) e^{iH_B t} (g_m b_m^\dagger + g_m^* b_m) e^{-iH_B t} \rho_B\} . \tag{35}$$

With $Z$ defined in (14), we have

$$C(t) = \frac{1}{Z} \sum_{n,m} g_n^* g_m \left\{ Tr_B \left( b_n^\dagger b_n e^{-\beta \omega_n b_n^\dagger b_n} \right) \left( e^{i\omega_n t} + e^{-i\omega_n t} \right) + e^{i\omega_n t} \right\} \delta_{nm} \tag{36}$$

$$= \sum_n |g_n|^2 \left\{ \left(e^{\beta\omega_n} - 1\right)^{-1} \left(e^{i\omega_n t} + e^{-i\omega_n t}\right) + e^{i\omega_n t} \right\} .$$

In the continuum limit of infinitely many bath modes, we introduce the density of states, $G(\omega)$, with the large-frequency cut-off at $\omega_c$. The summation over the modes is then replaced by the integration,

$$C(t) = \int d\omega J(\omega) \left[\cos(\omega t)\coth\left(\frac{\beta\omega}{2}\right) + i\sin(\omega t)\right] . \qquad (37)$$

where

$$J(\omega) \equiv G(\omega)|g(\omega)|^2 = \frac{\alpha}{2}\omega e^{-\omega/\omega_c} . \qquad (38)$$

In (38), we assumed the density of states in the standard Ohmic heat bath[25] form, with an exponential cut-off. In the notation of (31) and (32), we now get

$$\vec{K}(t) = 0 , \qquad (39)$$

$$\vec{R}(t) = \begin{pmatrix} -4\cos(Et)C_{\text{Re}}(t) & -E\delta(t) + 4\sin(Et)C_{\text{Re}}(t) & 0 \\ E\delta(t) - 4\sin(Et)C_{\text{Re}}(t) & -4\cos(Et)C_{\text{Re}}(t) & 0 \\ 0 & 0 & 0 \end{pmatrix} , \qquad (40)$$

$$C(t) = -\alpha(kT)^2 \operatorname{Re}\psi'\left(\frac{1 - i\omega_c t}{\omega_c/kT} + 1\right) + \frac{\alpha\omega_c^2}{2} \frac{1 - \omega_c^2 t^2}{\left(1 + \omega_c^2 t^2\right)^2} + i\alpha\omega_c^2 \frac{\omega_c t}{\left(1 + \omega_c^2 t^2\right)^2} , \qquad (41)$$

where $\psi'(z)$ is the derivative of the standard digamma function.

To simplify the calculations, we will consider the zero-temperature case from now on. In Appendix A, we demonstrate that the temperature-dependent term in (41) does not significantly contribute to the reduced density matrix for small temperatures. The Laplace transform of the real part of (41) becomes

$$C_{\text{Re}}(s) = F(s) , \qquad (42)$$

where we defined

$$F(s) = -\frac{\alpha s}{2}\left(\cos(s/\omega_c)\operatorname{ci}(s/\omega_c) + \sin(s/\omega_c)\operatorname{si}(s/\omega_c)\right) , \qquad (43)$$

which will be also considered in the complex-$s$ plane. Here $\operatorname{ci}(\varsigma)$ and $\operatorname{si}(\varsigma)$ denote the exponential cosine and sine functions. This explicit expression for $F(s)$ allows us to analyze the singularity structure of (34), which can be now rewritten, using (39), (40), and the notation $x_0 \equiv \xi_1(t=0)$, $y_0 \equiv \xi_2(t=0)$, $z_0 \equiv \xi_3(t=0)$, in the form

$$\xi_1(s) = \frac{x_0\left(2F(s-iE)+2F(s+iE)+s\right)+y_0\left(2F(s-iE)-2F(s+iE)-E\right)}{(s-iE+4F(s-iE))(s+iE+4F(s+iE))}, \qquad (44)$$

$$\xi_2(s) = \frac{y_0\left(2F(s-iE)+2F(s+iE)+s\right)-x_0\left(2F(s-iE)-2F(s+iE)-E\right)}{(s-iE+4F(s-iE))(s+iE+4F(s+iE))}, \qquad (45)$$

$$\xi_3(s) = \frac{z_0}{s}, \qquad (46)$$

where the components of the initial polarization vector satisfy $x_0^2 + y_0^2 + z_0^2 \leq 1$, with equality for the case of a pure initial spin state.

We note that the present approximation captures the property of the exactly solvable adiabatic decoherence model that $\xi_3(t) = z_0$, namely, that the diagonal elements of the density matrix remain unchanged for all times, $t \geq 0$. The Laplace transforms of the two other polarization vector components, $\xi_1(s)$ and $\xi_2(s)$, each have two branch-point singularities with superimposed poles, at $s = \pm iE$. As noted by Loss and DiVincenzo,[16] these singularities would be simple poles in the Markovian approximation. To calculate the inverse Laplace transform of (34), we integrate along the contour $S$ in the complex-$s$ plane; see Figure 1.

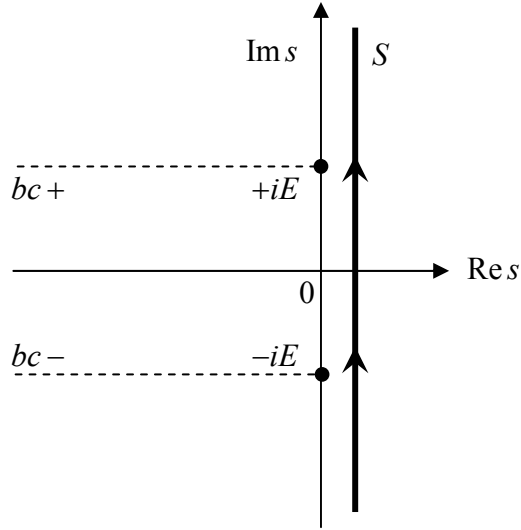

**Figure 1.** The integration contour and the singularity structure with two brunch-cuts, $bc+$ and $bc-$, in the complex plane of the Laplace-transformed time variable.

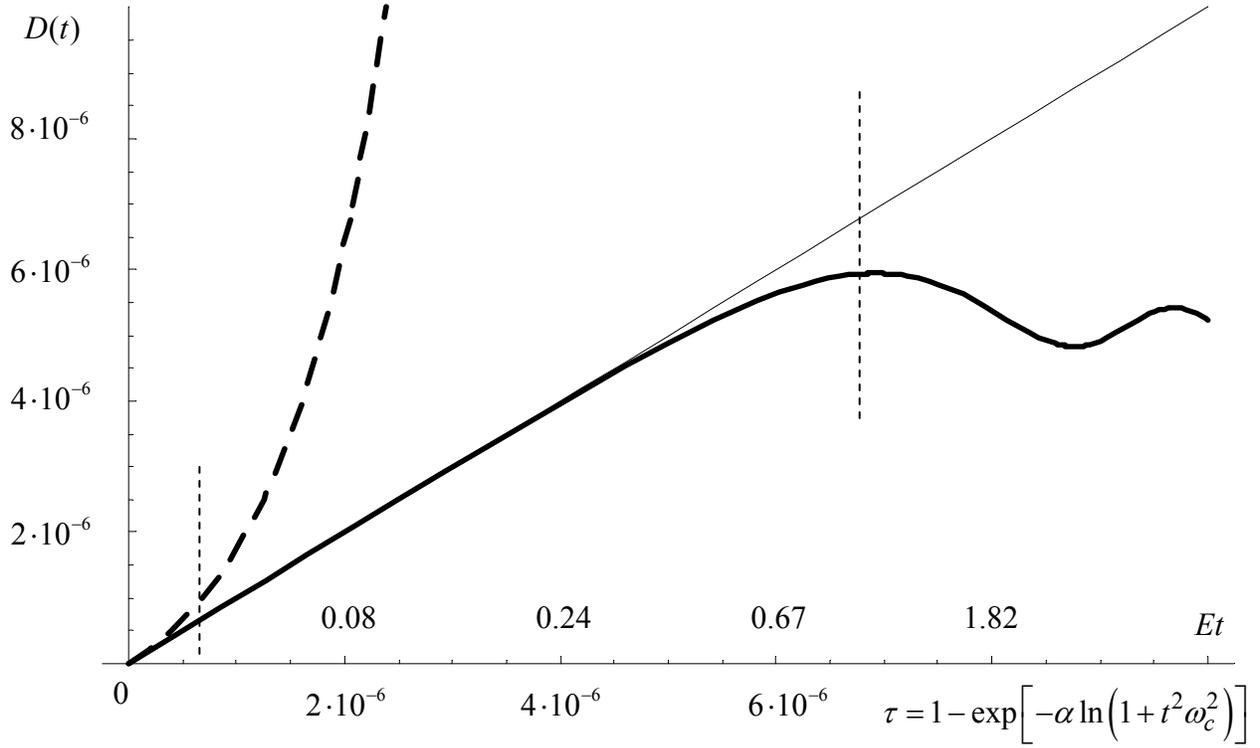

**Figure 2.** Decoherence measure for $\omega_c/E = 30$ and $\alpha = 10^{-6}$. The straight line for $D(\tau(t))$ gives the exact AD solution. The solid curve shows the NM-approximation results. The dashed curve is the SO-approximation. The vertical dashed lines correspond to $\omega_c t = 1$ and $Et = 1$.

To perform the integration, we move the contour to the left in the complex plane, to go around the two brunch cuts. The branch-cut integration can be carried out analytically to the first order in $\alpha$, which is consistent with the present perturbative approximation.[16] The reduced density matrix,

$$\rho(t) = \frac{1}{2}\begin{pmatrix} 1+\xi_3(t) & \xi_1(t)-i\xi_2(t) \\ \xi_1(t)+i\xi_2(t) & 1-\xi_3(t) \end{pmatrix}. \tag{47}$$

can be expressed via the function

$$f(t) \equiv e^{-2E/\omega_c}\Gamma(0, 2iEt - 2E/\omega_c) + e^{2E/\omega_c}\Gamma(0, 2iEt + 2E/\omega_c), \tag{48}$$

where $\Gamma(n,\varsigma)$ is the incomplete gamma function. We obtain the NM-approximation result,

$$\rho(t) = \frac{1}{2}\begin{pmatrix} 1+z_0 & (x_0-iy_0)e^{-iEt}\left(1+\alpha\left[f^*(t)-f^*(0)\right]\right) \\ (x_0+iy_0)e^{iEt}\left(1+\alpha\left[f(t)-f(0)\right]\right) & 1-z_0 \end{pmatrix}$$

(49)

$$= \frac{1}{2}\begin{pmatrix} 1+\xi_3(0) & (\xi_1(0)-i\xi_2(0))e^{-iEt}\left(1+\alpha\left[f^*(t)-f^*(0)\right]\right) \\ (\xi_1(0)+i\xi_2(0))e^{iEt}\left(1+\alpha\left[f(t)-f(0)\right]\right) & 1-\xi_3(0) \end{pmatrix}.$$

## 4. RESULTS AND DISCUSSION

To estimate the measure of decoherence $D(t)$ within the NM approximation, see (10), we calculate the eigenvalues of $\Delta\rho(t)$, which is the deviation from the ideal density matrix, obtained by setting $\alpha = 0$ in (49),

$$\lambda_\pm^{NM} = \pm\alpha|f(t)-f(0)|\sqrt{x_0^2+y_0^2}\ . \tag{50}$$

We thus get

$$D_{NM}(t) = \alpha|f(t)-f(0)|\ , \tag{51}$$

where $f(t)$ is given by (48). The AD-approximation, see (7) and (8), which happens to be the exact solution in our case, gives, for the Ohmic case,[15,19] equation (38),

$$B^2(t) = 4\alpha\int_0^\infty \frac{e^{-\omega/\omega_c}}{\omega}\sin^2\left(\frac{\omega t}{2}\right)\coth\left(\frac{\beta\omega}{2}\right)d\omega\ . \tag{52}$$

For the case of zero temperature, we get

$$B^2(t) = \alpha\ln\left(1+t^2\omega_c^2\right)\ . \tag{53}$$

For the adiabatic decoherence model, we then obtain the result

$$D_{AD}(t) = 1-e^{-\alpha\ln\left(1+t^2\omega_c^2\right)}\ . \tag{54}$$

We also note that the SO approximation gives

$$D_{SO}(t) = \alpha(t\omega_c)^2\ . \tag{55}$$

In Figure 2, the three decoherence measures are plotted for a typical choice of the parameter values. We observe that the NM approximation follows the exact AD-solution approximately up to the times of order $E^{-1}$, while the SO approximation breaks off for much shorter times, of order $\omega_c^{-1}$. Various estimates of the fault tolerant quantum computing[34-39] threshold require degree of decoherence not to exceed order $10^{-6}$-$10^{-4}$, for time scales of qubit "gate functions." Therefore, the ability to estimate decoherence for gate-function times $Et \sim 1$ is crucial. We find that the NM approximation[16] provides a reasonable estimate for the time scales of interest.

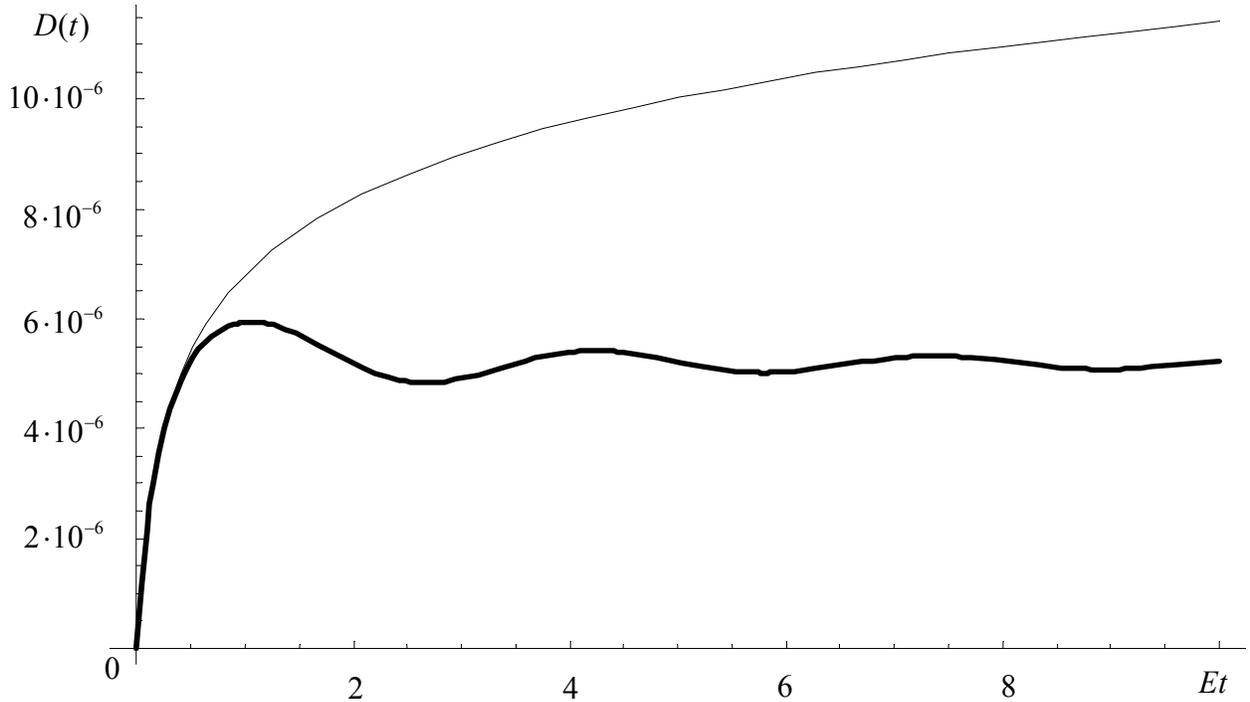

**Figure 3.** The NM (bottom curve) approximation and AD (top curve) exact result for intermediate times and beyond, for $\omega_c/E = 30$ and $\alpha = 10^{-6}$.

The calculated measure of decoherence within the NM approximation is shown for a larger time interval in Figure 3. As long as we consider the parameter regime $\omega_c/E \gg 1$, heuristically we find that the decoherence measure at $Et \sim 1$ is approximately $\sim 6\alpha$. We took $\alpha \sim 10^{-6}$ in Figures 2 and 3, to have the decoherence measure stay within the fault-tolerance[34-39] range, $10^{-6}$-$10^{-4}$. For a wide range of parameters, the qualitative behavior of $D_{NM}(t)$ remains the same. Past the times $Et \sim 1$, the measure $D_{NM}(t)$ develops oscillations and does not seem to approach any physically interesting value. While this is beyond the expected region of validity of the NM approximation, it is instructive to find the asymptotic behavior of $D_{\text{NM}}(t)$ at large times. Using the property of the incomplete gamma function that

$$\Gamma(\alpha,\varsigma) \approx \varsigma^{\alpha-1} e^{-\varsigma}, \qquad \text{for} \quad |\varsigma| \to \infty, \quad -3\pi/2 < \arg\varsigma < 3\pi/2, \tag{56}$$

we can simplify (48),

$$f(t \to \infty) \approx \frac{ie^{-2iEt}}{Et}. \tag{57}$$

We also use the property that

$$\Gamma(0, \varsigma \to 0) \approx -\ln\varsigma, \tag{58}$$

to obtain the large-time asymptotic behavior of (51),

$$D_{\text{NM}}(t \to \infty) \approx 2\alpha \left| \ln\left(\frac{2E}{\omega_c}\right) \right| + \alpha \frac{\sin 2Et}{Et}. \tag{59}$$

The second term of (59) shows the damped oscillations, while the first term gives the limiting value of the NM-approximation decoherence measure for large times. This analytical expression suggests an interesting observation that reduction of the degree of decoherence can be primarily achieved by a weaker coupling to the bath modes, the square of which is measured by $\alpha$, whereas other parameters only enter under the logarithm and have little effect on the value of $D(t)$.

In conclusion, we have considered short-time approximation techniques to estimate the rate of decoherence of the two-state quantum system weakly coupled to the environment. Comparing the results for the adiabatic model, we found that the SO approximation is valid only when $Et \ll 1$, whereas the NM approximation closely follows the AD approximation (which is exact in this case) up to $Et \sim 1$. For more general models, calculations within the NM and AD approximations, and comparison of the two results, are therefore expected to provide good estimates of the limit of validity of the short-time approximations and yield reasonable approximations of the reduced density matrix of the system up to intermediate time scales of quantum computing gate functions.

This research was supported by the National Security Agency and Advanced Research and Development Activity under Army Research Office contract DAAD-19-02-1-0035, and by the National Science Foundation, grant DMR-0121146.

# APPENDIX A

For small, but non-zero temperatures, we cannot omit the first term in (41). Here we will analyze the contribution of this temperature-dependent term to the NM-approximation density matrix and the calculated measure of decoherence. Our aim is to establish that the departure of the NM result, see Figure 3, from the exact curve, is not due to the zero-temperature assumption but rather due to the perturbative expansion in the strength of the interaction. The pattern of behavior illustrated in Figure 2 and 3 remains unchanged for small non-zero temperatures, though analytical expressions are more difficult to derive.

For $kT \ll \omega_c$, one can show that the leading correction in (43) is

$$F(s) = -\frac{\alpha s}{2}\left(\cos(s/\omega_c)\operatorname{ci}(s/\omega_c) + \sin(s/\omega_c)\operatorname{si}(s/\omega_c)\right)$$
$$- \alpha s\left(\cos(s/kT)\operatorname{Ci}(s/kT) + \sin(s/kT)\operatorname{si}(s/kT)\right) .$$
(A1)

The added term has the same structure as the zero-$T$ term. The integration along the contour $S$, see Figure 1, can be carried out as before. To the first order in $\alpha$, we obtain

$$f(t) = f_T(t) = f_{T=0}(t) + \delta f_T(t) ,$$
(A2)

where

$$\delta f_T(t) = e^{-2E/kT}\Gamma(0, 2iEt - 2E/kT) + e^{2E/kT}\Gamma(0, 2iEt + 2E/kT) .$$
(A3)

This will lead to an additional term in the density matrix, $\rho(t,T) = \rho(t,0) + \delta\rho(t,T)$, cf. (49), where

$$\delta\rho(t,T) = \frac{1}{2}\begin{pmatrix} 0 & (x_0 - iy_0)e^{-iEt}2\alpha\left(\delta f_T^*(t) - \delta f_T^*(0)\right) \\ (x_0 + iy_0)e^{iEt}2\alpha\left(\delta f_T(t) - \delta f_T(0)\right) & 0 \end{pmatrix} .$$
(A4)

The following approximate estimate can be derived from (A3),

$$\max_t \left|\delta f_T(t) - \delta f_T(0)\right| \lesssim \begin{cases} 2kT/E & \text{for } kT/E < 1 \\ 2\sqrt{kT/E} & \text{for } kT/E > 1 \end{cases} .$$
(A5)

The measure of decoherence $D(t,T)$, see (10) and (11), will be modified from its zero-$T$ value by the correction, $\delta D_T(t)$. For the important regime of validity of the NM approximation, $kT/E < 1$, one can obtain an approximate estimate

$$\max_t \left|\delta D_T(t)\right| < 4\alpha kT/E .$$
(A6)

This establishes that the temperature-dependent correction does not significantly contribute the measure of decoherence $D(t,T)$ in the NM approach, provided the temperature is small.